# Quantum critical behaviour in $Na_xCoO_2$

F. Rivadulla[1*], M. Bañobre-López[1], M. A. López-Quintela[1] & J. Rivas[2].

[1]*Physical-Chemistry* and [2]*Applied Physics Department, University of Santiago de Compostela, 15782-Santiago de Compostela, Spain.*

(*Corresponding author (F.R.) e-mail: qffran@usc.es)

**Abstract.-** We report the divergence of the nonlinear component of the bulk susceptibility in $Na_xCoO_2$ ($0.3<x<0.62$) as $T\rightarrow 0K$. These experiments provide an striking evidence of the existence of a ferromagnetic phase transition at zero Kelvin (quantum phase transition). The possible role of magnetic fluctuations in the superconductivity is discussed to the light of the observed (H,T) scaling of the magnetization, which implies a local character of the fluctuations.

The breakdown of Fermi-Liquid theory in the paramagnetic phase of itinerant ferromagnets[1,2] and elemental paramagnets[3], as well as unconventional forms of magnetically mediated superconductivity in heavy fermions[4,5], have been related to the proximity to a quantum critical point (QCP). In 2D, the magnetic excitations at the critical point depart from the traditional prediction of a spatially delocalized wave, becoming localized at atomic length scales[6]. The suppression of the long-wavelength fluctuations of the order parameter was predicted to have deep consequences on many observable physical properties of strongly correlated 2D metals close to a QCP. Here we report the critical behaviour of the nonlinear susceptibility, $\chi_3(T\rightarrow 0\ K)$ in a 2D paramagnetic metal: $Na_xCoO_2$. The divergence of $\chi_3(T \rightarrow 0\ K)$ provides a striking evidence of the occurrence of a ferromagnetic QCP in this material, in which the static spin susceptibility satisfies a series of scaling laws predicted for a local-moment QCP. An unconventional form of superconductivity in $Na_xCoO_2$[7,8] together with the localized character of the magnetic fluctuations, presents a new challenge for the understanding of magnetically meditated superconductivity and its relationship to the proximity to a magnetic QCP.

Much of the original interest in $Na_{0.5}CoO_2$ was triggered by the unusual combination of large thermoelectric power and low metallic resistivity in this compound[9], which opened a completely unexpected path to new thermoelectric materials. But it was the report of a crossover in the effective electronic dimensionality above a certain temperature[10] and in particular, the finding of superconductivity below 5 K in $Na_xCoO_2 \cdot yH_2O$ (x ≈ 0.35) what drove a vast amount of the research towards the outstanding fundamental properties of this family of cobalt oxides.

In $Na_xCoO_2$, the $CoO_2$ planes adopt a 2D hexagonal symmetry, conforming the first example of a superconducting system with such a geometry. The proximity to a localized charge-ordered phase[11] makes it very tempting a straightforward comparison with the cuprates, in terms of a hole-doped charge-ordered triangular lattice. However, $^{59}Co$ NMR and NQR experiments[8,12] revealed an unconventional form of a superconducting spin-triplet phase, probably related to the existence of magnetic spin fluctuations in the metallic counterpart. On the other hand, LDA calculations[13] where found to predict a FM ground state for $Na_xCoO_2$ in a broad range of x down to 0.3. This discrepancy with the experiment was tentatively attributed to the proximity to a quantum phase transition (QPT), in analogy to other materials like $Sr_3Ru_2O_7$, also wrongly predicted to be in the ordered state.

By investigation of the field and temperature dependence of the bulk magnetization we demonstrate here that metallic $Na_xCoO_2$ is in fact close to a zero-temperature, second order, FM phase transition in a wide compositional range (0.62<x<0.3, except maybe around x=0.5 where a charge ordering develops). At low temperatures, typically below 10 K, there is a continuous negative departure from the linear M(H) behavior expected in a conventional metallic paramagnet (PM) (Fig.1), although no magnetic order was detected in the susceptibility, resistivity, etc. This deviation is even more accused for the metallic precursors of the superconducting phases with optimal $T_C$ (x≈0.3). In any case, no metamagnetism was induced up to 9T for any composition.

It is at the vicinity of a phase transition where non-linear effects become more relevant and the magnetization must be treated as an expansion of the magnetic field. In the paramagnetic range M is an odd function of H

$$M = \chi_1 H + \chi_3 H^3 + \chi_5 H^5 + \ldots \qquad [1]$$

where the third-order susceptibility, $\chi_3$, is normally quoted as the nonlinear susceptibility. Its critical behavior is commonly used to characterize the spin-glass transition[14] but it has been rarely applied for the characterization of conventional second order FM-PM phase transitions. In this case, a mean field treatment gives a negative divergence of $\chi_3$ above the Curie point, while below the transition temperature $\chi_3$ diverges positively as $T \rightarrow T_C$[15]. This provides an accurate method to determine the critical point, *i.e.* $T_C$. We have fitted the M(H) curves at each temperature to extract $\chi_3$ and the results are plotted in Fig. 2. There is a clear departure from the high temperature value ($\chi_3 \rightarrow 0$) and a negative divergence of of $\chi_3$ below $\approx$ 10K. The nonlinear susceptibility shows the typical signatures of a system just above the FM-to-PM transition temperature (completely different behavior would be expected in an antiferromagnetic (AF) to PM transition).

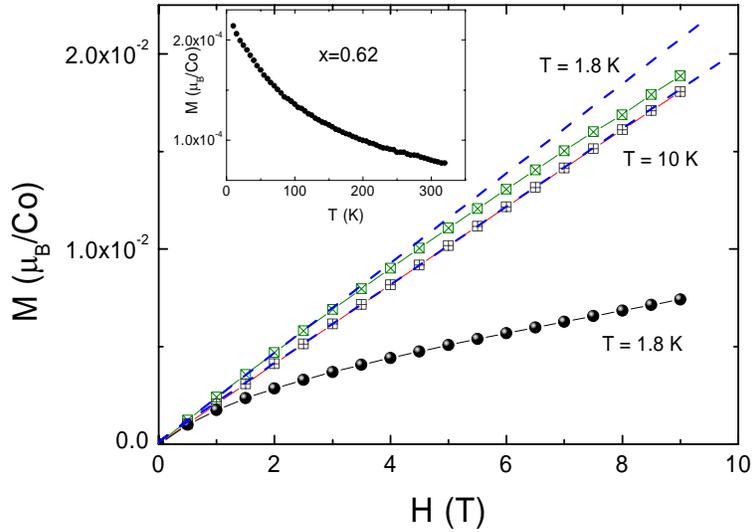

**Fig. 1**. Magnetization vs. applied field at two different temperatures in polycrystalline $Na_xCoO_2$, (x=0.62, open symbols, and x=0.3 closed symbols) up to 9T and down to 1.8 K. The departure from the linear behaviour is clear in x=0.62 above $\approx$3T at 1.8K. The effect is even more evident in x=0.3. The dashed straight lines were generated from the initial susceptibility. *Inset*: Temperature dependence of the magnetization at 1 kOe. The expected metallic independent value is not followed at all, but a modified Curie-Weiss law.

The divergence of $\chi_3$ is a clear indication of the existence of a FM singularity and implies a divergence of the spin correlation length, although the transition temperature was not reached even at 1.8 K (the lowest temperature probed in this work). $\chi_3$ does not show any sign of rounding still at 1.8 K, which indicates that we are still far from the transition temperature. This behavior of the nonlinear susceptibility provides a clear and strong proof of the existence of a zero Kelvin FM phase transition in $Na_xCoO_2$.

In the conventional theory of metals close to a magnetic QCP, long-wavelength fluctuations of the order parameter (paramagnons) are the relevant magnetic excitations. On the other hand, for 2D magnetic systems Si *et al.*[6] predicted the existence of a different kind of magnetic QCP, in which the magnetic fluctuations are spatially localized. $Na_xCoO_2$ is a strongly anisotropic system[10], in which 2D magnetic fluctuations are expected, making this a good candidate for the occurrence of a local QCP. This scenario of a locally critical quantum phase transition (QPT) provides a series of scaling equations for the bulk susceptibility in the (H, T) planes, not predicted when all the critical degrees of freedom are spatially extended. Two easily verifiable laws, derived from the modified Curie-Weiss form of the spin susceptibility in this scenario are[6,16]

$$\chi^{-1}(T) \propto T^{\alpha} \qquad [2]$$

$$\chi T^{-\alpha} \propto \left(H/T^{\beta}\right) \qquad [3]$$

The scaling derived from eq.[2] and [3] is checked in Fig.2 (inset) and 3. The samples follow the modified Curie-Weiss law at the proximity of the QCP for a broad range of compositions down to at least x≈0.3 (inset of Fig. 2). Best fits were obtained for a value of the exponent α=0.80(2). It must be noted that the slope and the intercept changes with the Na content, reflecting a change in this Curie constant and Weiss temperature. In Figure 3 we show the collapse of the magnetization isotherms suggested by eq.[3], with a exponent α slightly lower than that obtained from the plot of equation [2]. Our analysis, throughout the verification of the scaling laws, is a strong argument to support the local character of the magnetic fluctuations close to the QCP. The existence of unconventional local spin excitations is the cause not only of the scaling but can be also the source of important departures from the standard Fermi liquid model. In fact, Rivadulla *et al.*[17] reported a non-Fermi liquid $T^{3/2}$ temperature dependence of the *ab*-plane resistivity in single crystals of this material (see also the inset of fig.1 for non-metallic dependence of the susceptibility). The 3/2 exponent is recurrent in other systems close to a magnetic phase transition at 0 K, like MnSi[1] or the β phase of pure Mn[3], regardless of the local or extended character of the spin fluctuations close to the QCP. A conclusive explanation about the origin of this exponent is still missing, although it seems clear now that it is more general than previously anticipated.

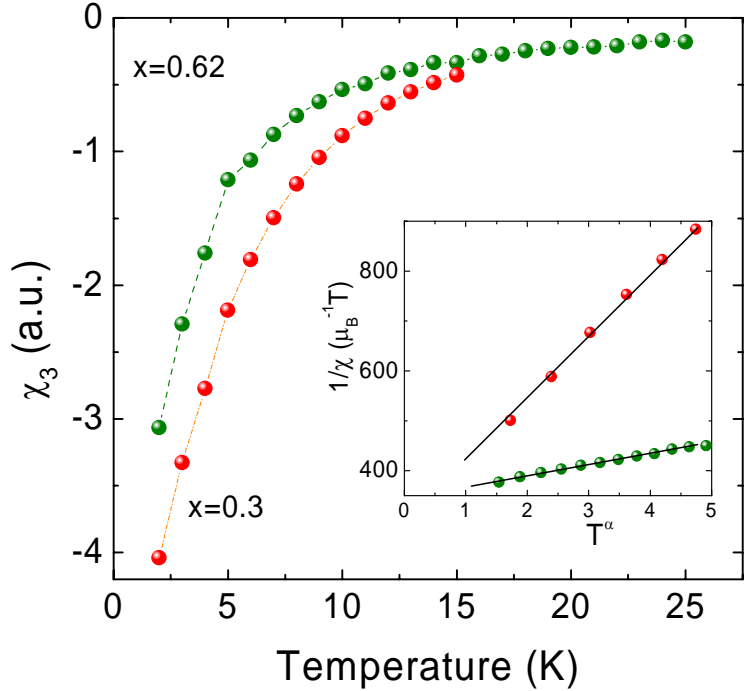

**Fig. 2**. Divergence of the nonlinear susceptibility as T→0 K for x≈0.62 (circles) and x≈0.3 (squares). Similar behaviour was observed in the whole range of x checked (between 0.3 and 0.62), even in the superconducting samples. *Inset*: Power-law dependence of the inverse susceptibility given by eq.[2]. Best linear fits were obtained for α=0.80(2).

As we already mentioned, the report of superconductivity in hydrated phases of $Na_xCoO_2$ created an enormous experimental and theoretical interest, in part due to the role that the geometric/magnetic properties of the triangular cobalt-oxygen lattice could play for a general understanding of the superconductivity phenomenon. We have found that the divergence of $\chi_3$, and hence of the amplitude of the FM fluctuations, is also present for superconducting samples, and is very clear for the range of doping when the $T_C$ is optimal[18]. Through $^{59}$Co NMR, Ishida *et al*.[8] and Waki *et al*.[12] demonstrated that superconductivity in $Na_xCoO_2$ is characterized as spin-triplet state with p-wave symmetry, and the nuclear spin lattice relaxation time in the superconducting samples is consistent with the existence of low-energy FM fluctuations. From these results an important role was attributed to the spin fluctuations in the unconventional superconductivity, in a similar fashion to other materials like $Sr_2RuO_4$[19] or $UGe_2$[4]. However, we would like to stress out that to the light of some of our results, this comparison is not trivial.

In the traditional theory of metals close to a QCP, correlations become infinitely long range in space and time as the critical point is approached, and the fundamental assumption of the Fermi-liquid model, *i.e.* the short-range character of the quasiparticles correlations, is violated. At this point many (or all) of the observable consequences of the model (linear specific heat, weakly-T dependent susceptibility, $T^2$ dependence of the resistivity at low temperature, etc.) are expected to fail. But there is another important consequence of the long-range character of the magnetic waves: there is an effective magnetic interaction that can be attractive for fermions with the same spin orientation. This interaction favours the formation of Cooper pairs in the triplet state and p-wave orbital symmetry, and under certain conditions can lead to a magnetically mediated superconductivity.

This seems to be the case for the itinerant ferromagnet $UGe_2$ described by S. S. Saxena *et al*[4]. where magnetically mediated superconductivity was observed below $\approx$ 1 K in a narrow range of pressures near the critical value at which the FM transition temperature is tuned down to 0 K. This kind of anisotropic superconductivity is easily disestablished by scattering processes and is expected to be observed only at very low temperatures, when superconducting correlation length $\xi$ is much smaller than the electronic mean-free path $\lambda$. Although at this moment there are no good enough data of $\xi$ and $\lambda$ in the same specimen of $Na_xCoO_2$ to make an accurate estimation, the low residual resistivity measured in single crystals of $Na_xCoO_2$ in the *ab*-plane ($\approx$8 $\mu\Omega$cm)[17] corresponds to a $\lambda$ of several hundreds of Å, which makes highly probable that this material is at the clean limit ($\xi < \lambda$). From this point of view, the achievement of the clean limit condition together with our report of proximity to a FM-QCP, will make $Na_xCoO_2$ a good candidate for the occurrence of triplet superconductivity mediated by FM-spin fluctuations.

But however, the scaling laws we have probed to hold in the critical region of $Na_xCoO_2$, show that the magnetic fluctuations close to the QCP are spatially localized, and do not form a magnetic wave that propagates over a long range. In principle, this is a prerequisite for the occurrence of magnetically mediated superconductivity, and maybe the cause of the absence of superconductivity in doped metals in which scaling was observed, like in $CeCu_{6-x}Au_x$[16]. This imposes an important restriction to a straightforward applicability of the existing theories of superconductivity at the verge of a FM-QPT to the case of $Na_xCoO_2$. The strength of the superconducting state against oxygen vacancies and the relative high values of $T_C$, are not easily understood either with existing models

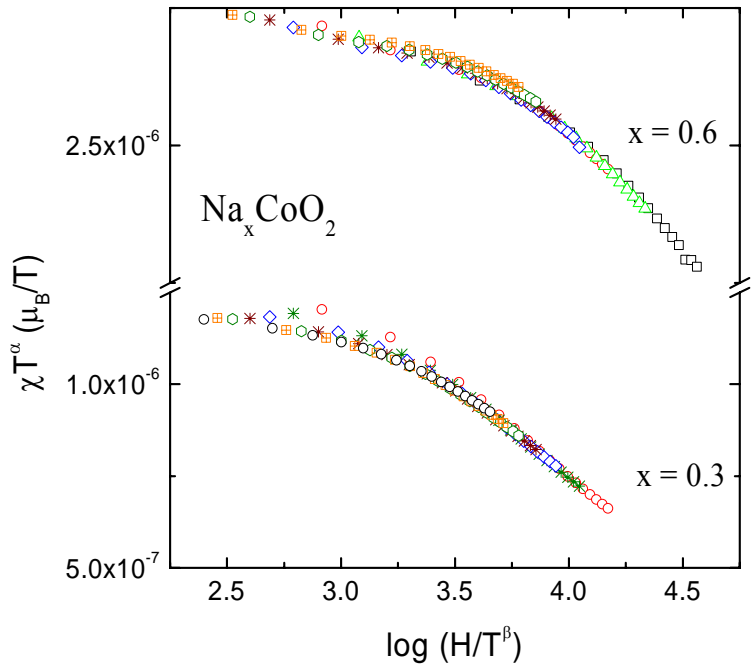

**Fig. 3**. Scaling plot of the susceptibility in the critical region above $T_C$, following eq.[3].

Therefore, an anomalous kind of magnetically meditated superconductivity in $Na_xCoO_2$ is in contradiction with the local character of the magnetic fluctuations close to the QPT. One possibility is that the superconductivity and the magnetism arise from two different channels, in a scenario in which delocalized waves and localized fluctuations coexist in the same specimen. Whether this is the case or if on the other hand there is room for magnetically mediated superconductivity into the locally critical QPT theory, these experiments present an important challenge for current models of magnetic coupling and quantum criticality.

In summary, we have demonstrated the proximity of $Na_xCoO_2$ to a FM, second-order phase transition at 0 K. The local character of the magnetic fluctuations imposes an important restriction to the applicability of current models of magnetically mediated superconductivity. These experiments provide an scenario in which many of the anomalies observed in this compound naturally emerge, but also constitute an important challenge to our understanding of the relationship between magnetic superconductivity and quantum criticality.

**Acknowledgments.-** We want to acknowledge Dr. L. E. Hueso and Dr. S. S. Saxena for useful comments and critical reading of the manuscript. F.R. acknowledges financial support from the MC&T of Spain through the program Ramón y Cajal. This work was supported by DGI/MC&T of Spain through project FEDER MAT2001-3749.